\documentclass[%
 reprint,
superscriptaddress,
 amsmath,amssymb,
 aps,
 prl,
]{revtex4-2}

\usepackage{graphicx}
\usepackage{dcolumn}
\usepackage{bm}
\usepackage{hyperref}
\usepackage{amsmath}

\begin{document}


\title{Orientational melting in a mesoscopic system of charged particles}

\author{Lucia Duca}
\affiliation{Istituto Nazionale di Ricerca Metrologica (INRiM), 10135 Torino, Italy}
\affiliation{European Laboratory for Nonlinear Spectroscopy (LENS), 50019 Sesto Fiorentino, Italy}
\author{Naoto Mizukami}
\affiliation{Istituto Nazionale di Ricerca Metrologica (INRiM), 10135 Torino, Italy}
\affiliation{European Laboratory for Nonlinear Spectroscopy (LENS), 50019 Sesto Fiorentino, Italy}
\affiliation{Politecnico di Torino, 10129 Torino, Italy}
\author{Elia Perego}
\altaffiliation{Current address: Department of Physics, University of California, Berkeley, CA, USA}
\affiliation{Istituto Nazionale di Ricerca Metrologica (INRiM), 10135 Torino, Italy}
\affiliation{European Laboratory for Nonlinear Spectroscopy (LENS), 50019 Sesto Fiorentino, Italy}
\author{Massimo Inguscio}
\affiliation{European Laboratory for Nonlinear Spectroscopy (LENS), 50019 Sesto Fiorentino, Italy}
\affiliation{Istituto Nazionale di Ottica del Consiglio Nazionale delle Ricerche (CNR-INO), 50019 Sesto Fiorentino, Italy}
\affiliation{Department of Engineering, Campus Bio-Medico University of Rome, 00128 Rome, Italy}
\author{Carlo Sias}
\email{c.sias@inrim.it}
\affiliation{Istituto Nazionale di Ricerca Metrologica (INRiM), 10135 Torino, Italy}
\affiliation{European Laboratory for Nonlinear Spectroscopy (LENS), 50019 Sesto Fiorentino, Italy}
\affiliation{Istituto Nazionale di Ottica del Consiglio Nazionale delle Ricerche (CNR-INO), 50019 Sesto Fiorentino, Italy}


\begin{abstract}
A mesoscopic system of a few particles exhibits behaviors that strongly differ from those of a macroscopic system. While in a macroscopic system phase transitions are universal, a change in the state of a mesoscopic system depends on its specific properties \cite{Hill1962}, like the number of particles, to the point that changes of state can be disfavored for specific “magic numbers” \cite{Haberland2005}.
A transition that has no counterpart in the macroscopic world is orientational melting, in which localized particles with long-range repulsive interactions forming a two-dimensional crystal become delocalized in common circular or elliptical trajectories.
Orientational melting has been studied extensively with computer simulations \cite{Filinov2001,Pupillo2010,Boning2008,Bedanov1994,Schweigert2000} and witnessed in a few pioneering experiments \cite{Bubeck1999,Juan1998,Wineland1987,Melzer2012}. However, a detailed experimental investigation fully revealing its non-universal nature has been missing so far.
Here we report the observation of orientational melting in a two-dimensional ensemble of up to 15 ions with repulsive Coulomb interaction. 
We quantitatively characterize orientational melting, and compare the results with a Monte Carlo simulation to extract the particles’ kinetic energy. We demonstrate the existence of magic numbers \cite{Schweigert1995}, and control locally the occurrence of melting by adding a pinning impurity.
Our system realizes a fully-controllable experimental testbed for studying the thermodynamics of small systems \cite{Dauxois2002}, and our results pave the way for the study of quantum phenomena in systems of delocalized ions, from the emergence of quantum fluctuations \cite{Bonitz2008B} and quantum statistics \cite{Roos2017}, to the control of multi-shell quantum rotors \cite{Urban2019}.
\end{abstract}
\maketitle

\subsection{Introduction}
A system of confined particles with long-range repulsive interactions undergoes crystallization at a sufficiently low temperature, i.e. the particles become localized in a self-ordered structure. Crystallization has been observed in several mesoscopic physical systems, including electrons in quantum dots \cite{Reimann2002}, trapped ions \cite{Diedrich1987,Wineland1987}, liquid Helium \cite{Grimes1979} and atomic clusters \cite{Schauss2012}.
When confined in an isotropic two-dimensional (2D) potential, a mesoscopic crystal of a few particles can melt in the angular degree of freedom since there is no preferential orientation of the crystal. This orientational melting is triggered by thermal or quantum fluctuations \cite{Bonitz2008B}, and results in a delocalization of the particles in concentric circular trajectories (shells), while the system remains localized radially \cite{Filinov2001, Boning2008}.
When undergoing orientational melting, the particles change their state in a process that resembles a phase transition for a macroscopic system, but out of the thermodynamic limit. Therefore, orientational melting is a non-universal phenomenon, i.e. it occurs at conditions that strongly depend from the specific properties of the system. 
Orientational melting has been studied with computer simulations with several types of long-range interactions, from Yukawa potential \cite{Totsuji2001} to Coulomb \cite{Bedanov1994, Belousov_1999} or dipolar interactions \cite{Belousova2000, Pupillo2010}. 
One of the main findings of these studies is that adding or removing a single particle can result in dramatically different collective properties, as expected from the non-universality of the phenomenon. In particular, there exist special "magic numbers" of particles for which orientational melting is particularly disfavoured \cite{Schweigert1995}. Despite of the vast theoretical literature, the few experimental observations reported so far \cite{Bubeck1999,Juan1998,Wineland1987,Melzer2012} could not reach the high level of control of the experimental parameters that is needed to fully capture the non-universality features of orientational melting.

Here, we directly observe and characterize orientational melting in a two-dimensional crystal of Ba$^+$ trapped ions. The main advantages of using trapped ions are the possibility of precisely setting the number of particles and of creating two-dimensional crystals by using external electric fields \cite{Mitchell1998,Richerme2021,Okada2010,Li2017}. Moreover, we can observe the occurrence of the transition in real time by using fluorescence imaging.
We demonstrate the non-universality of orientational melting and quantitatively characterize its occurrence by measuring angular density-density correlations and the angular spread of the single ion density distribution.
We observe that orientational melting occurs under conditions that strongly depend on the number of particles, and find excellent agreement with the results of a Monte Carlo simulation. Moreover, we are able to locally inhibit melting by adding a single impurity with a different mass. Interestingly, in a sufficiently large ensemble, the presence of a pinning impurity leads to the creation of a unique structure in which localized and delocalized particles co-exist.
Our system provides a testbed for exploring the thermodynamics of mesoscopic systems \cite{Hill1962,Dauxois2002}, and the full control on the experimental parameters that we demonstrate paves the way to accessing new groundbreaking quantum regimes for delocalized strongly-interacting particles \cite{Bonitz2008B,Roos2017,Urban2019}.
\subsection{Orientational melting of a two-dimensional crystal}

\begin{figure}[t]
    \centering
    \includegraphics[width=0.98\columnwidth]{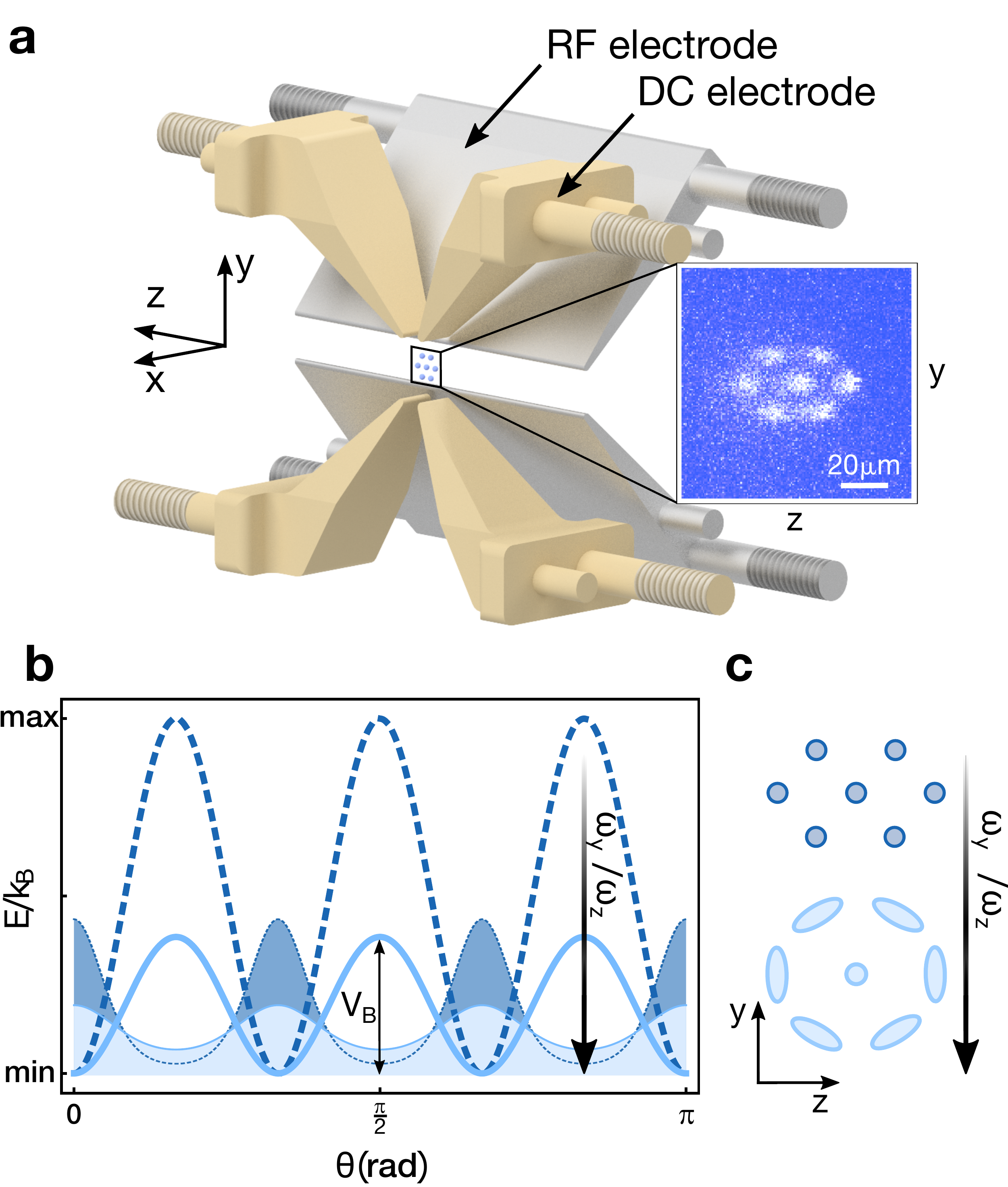}
    \caption{\textbf{Sketch of the physical system.} \textbf{a}, The ion trap is composed by 4 RF (gray, only two shown) and 4 DC (yellow) electrodes. The trap frequency ratio of the weakly confined directions ($\omega_y/\omega_z$) is controlled by changing the applied electrical potentials (see Supplementary Materials). The inset shows a picture of a two-dimensional crystal of 7 $^{138}$Ba$^+$ ions in a potential with trap frequencies $(\omega_x,\omega_y,\omega_z)=(400,121,97)$~kHz. \textbf{b}, A reduction of the $\omega_y/\omega_z$ ratio towards unity corresponds to a decrease of the height $V_B$ of the potential barrier associated to the rigid rotation of the crystal. When $V_B$ is reduced (e.g. from dark to light blue in the figure) the particles spatial distribution spreads, as illustrated in the figure by the two shaded thermal distributions and their corresponding sketches in \textbf{c}. When the barrier is further lowered, the particles delocalize along a closed trajectory and the crystal undergoes orientational melting.
   }
    \label{fig:1}
\end{figure}

The Hamiltonian that describes $N$ singly-charged particles in a two-dimensional harmonic potential is: 
\begin{equation}
H=\sum_{i=1}^N \left(\frac{\mathbf{p_i}^2}{2 m_i}+ \frac{m_i}{2} \left(\omega_y^2 y_i^2 +\omega_z ^2 z_i^2\right)  
+ \sum_{j>i}^N \frac{\alpha}{|\mathbf{r_i}-\mathbf{r_j}|}\right)
\label{Hamiltonian}
\end{equation}
where $N$ is the total number of ions, $m_i$ is the ion mass, $\mathbf{r_i}=(y_i,z_i)$ is the position of the i-th ion in the two-dimensional $y$-$z$ plane with trap frequencies $\omega_y$ and $\omega_z$, $\mathbf{p_i}$ is the i-th ion momentum, $\alpha=e^2/(4\pi \epsilon_0)$, $e$ is the electron charge, and $\epsilon_0$ is the vacuum permittivity. 
The Hamiltonian in Eq.\ \ref{Hamiltonian} is well approximated by the pseudo-potential created by the Paul trap shown in Fig.\ \ref{fig:1}. So far we are not considering the presence of micromotion, whose effects result in a minor correction (see Supplementary Materials). 
The trap frequencies depend on the voltages applied to the RF and DC electrodes, $V_{RF}$ and $V_{DC}$ respectively, and can be expressed in terms of the Mathieu parameters $\mathbf{a}$ and $\mathbf{q}$ \cite{Leibfried2003}.
Crucially, we can continuously change the ratio $\omega_y/\omega_z$ by varying only the parameter $V_{DC}$ while keeping the dynamics in two dimensions, i.e. $\omega_x\gg\omega_y,\omega_z$ \cite{Perego2020} (see Supplementary Materials). In this trap, the Doppler cooled ions self-arrange in a 2D crystal with elliptical shape, as shown in Fig.\ \ref{fig:1}\textbf{a}. In the image, the ions have a preferred orientation that originates from the anisotropy of the trap.
Fig.\ \ref{fig:1}\textbf{b} shows the energy of the system as the crystal is rigidly rotated by an angle $\theta$ from its equilibrium position. This energy, which we evaluate with a Monte Carlo simulation, has for $\omega_y\simeq\omega_z$ a sinusoidal shape with amplitude $V_B/2$ \cite{Schweigert1995} (see Supplementary Materials). The ions are localized when the angular potential barrier $V_B$ is much higher than the ions kinetic energy $E_T$. 
The potential barrier $V_B$ can be controlled by changing the ratio $\omega_y/\omega_z$, and when $V_B$ becomes comparable to the kinetic energy the ions' angular distribution starts spreading (see Fig.\ \ref{fig:1}\textbf{b-c}), i.e. the crystal starts melting.

\begin{figure}[t]
    \centering
    \includegraphics[width=\columnwidth]{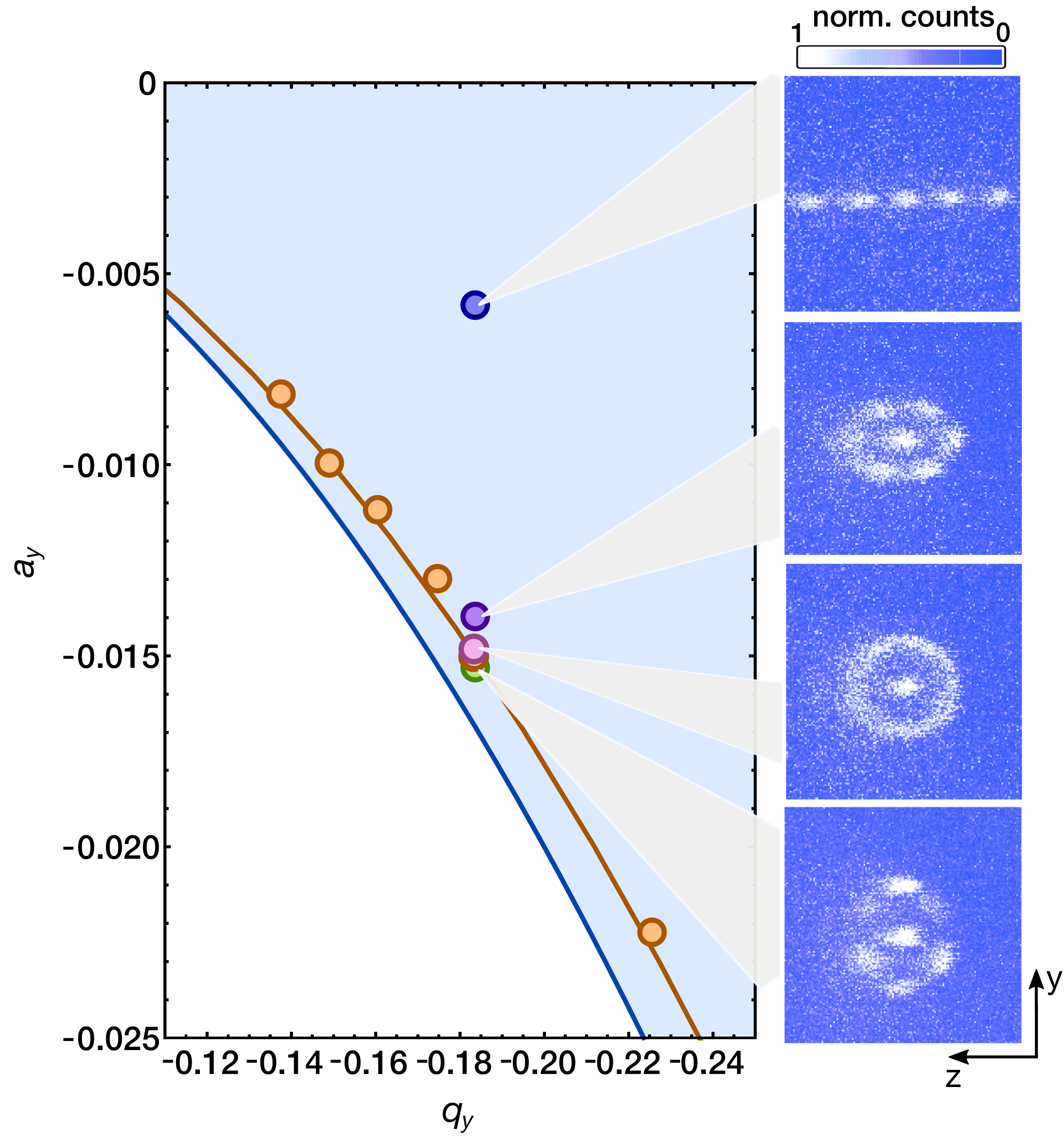}
    \caption{\textbf{Accessing orientational melting by changing the particles' confining potential.} Left: stability diagram of the ion trap calculated for $^{138}$Ba$^+$, expressed as a function of $a_y$ and $q_y$ (see Supplementary Materials). Each combination of Mathieu parameters $\mathbf{a}$ and $\mathbf{q}$ within the stability region (blue area) correspond to different trap frequencies. The yellow curve corresponds to the condition $\omega_y/\omega_z=1$, as expected from the simulation of our trap. The location of this curve is confirmed by experimental data (yellow dots) obtained from fitting the ions' spatial distribution and corresponding to a radially symmetric crystal (see Supplementary Materials). 
    Right: images of 5 (i) and 7 (ii-iv) $^{138}$Ba$^+$ ions at $q_y=-0.182$ and different values of $a_y$. The images illustrate how crystallization and ellipticity change as a function of $a_y$ across the melting transition. The images are taken at $\omega_y/\omega_z=(3.9, 1.2, 1.1, 0.9)$, $\omega_y=2\pi\times(246, 121, 107, 91)$~kHz and $q_y=-0.182$, top to bottom.}
    \label{fig:2}
\end{figure}

In the experiment, we access the melting transition by varying $\bf{a}$ and $\bf{q}$ within the stability diagram, as illustrated in Fig.\ \ref{fig:2}. 
The experimental images of Fig.\ \ref{fig:2} are taken with 5 to 7 trapped ions and illustrate how the variation of the trap parameter $\bf{a}$ affects the aspect ratio of the ion crystal and the angular localization of the trapped particles.
As $|\bf{a}|$ is increased, the ion crystal shape changes from a line to an ellipse (i-ii). When $\omega_y/\omega_z = 1$, we observe orientational melting as the ions are completely delocalized along a circular trajectory (iii). Importantly, the ion crystal recovers when $|\bf{a}|$ is further increased (iv). This is a clear indication that the loss of crystallization reflects a change of state of the system and not a trivial effect caused either by instabilities arising at the edge of the stability diagram \cite{Emmert1993} or by the micromotion increase occurring when the ions lie out of the trap z-axis \cite{Berkeland1998}.
We note that melting occurs for all the pairs of parameters $(\bf{a},\bf{q})$ for which $\omega_y/\omega_z = 1$. These values are represented in Fig.\ \ref{fig:2} with the yellow line, which lies within the trap stability diagram. The curve, which was calculated from the theoretical model of the trap, is in excellent agreement with the experimental data (yellow circles) corresponding to the observation of circular trajectories (see Supplementary Materials).

\subsection{Characterization of orientational melting}

To better characterize the onset of melting, we analyze images taken with different ion numbers, fixed $q_y=-0.182$, and different trap ratios $\omega_y/\omega_z$ in proximity of the melting transition.  
Each image records the fluorescence light, and therefore provides a spatial density distribution of the particles over the exposure time. We quantify the loss of angular ordering of the ions by using the angular correlation function
\begin{equation}
g(\Delta \theta) =\frac{\sum_{\theta = 0}^{2 \pi} n(\theta) n(\theta + \Delta \theta) - \sum_{\theta = 0}^{2 \pi} n(\theta)^{2}}{\sum_{\theta = 0}^{2 \pi} n(\theta)^{2}},\end{equation} 
which reflects the probability of finding two particles at an angular distance $\Delta \theta$ along an elliptical trajectory enclosing the positions of the ions around their center of mass (see Supplementary Materials). Notably, if the ions are forming a crystal, $g(\Delta\theta)$ will show a modulation with period $\theta_{NT}=2\pi/N_{T}$, where $N_{T}$ is the number of ions in the elliptical path. We calculate the amplitude $C$ of this modulation and use it to quantify the degree of localization of the ions. 

\begin{figure}
    \centering
    \includegraphics[width=\columnwidth]{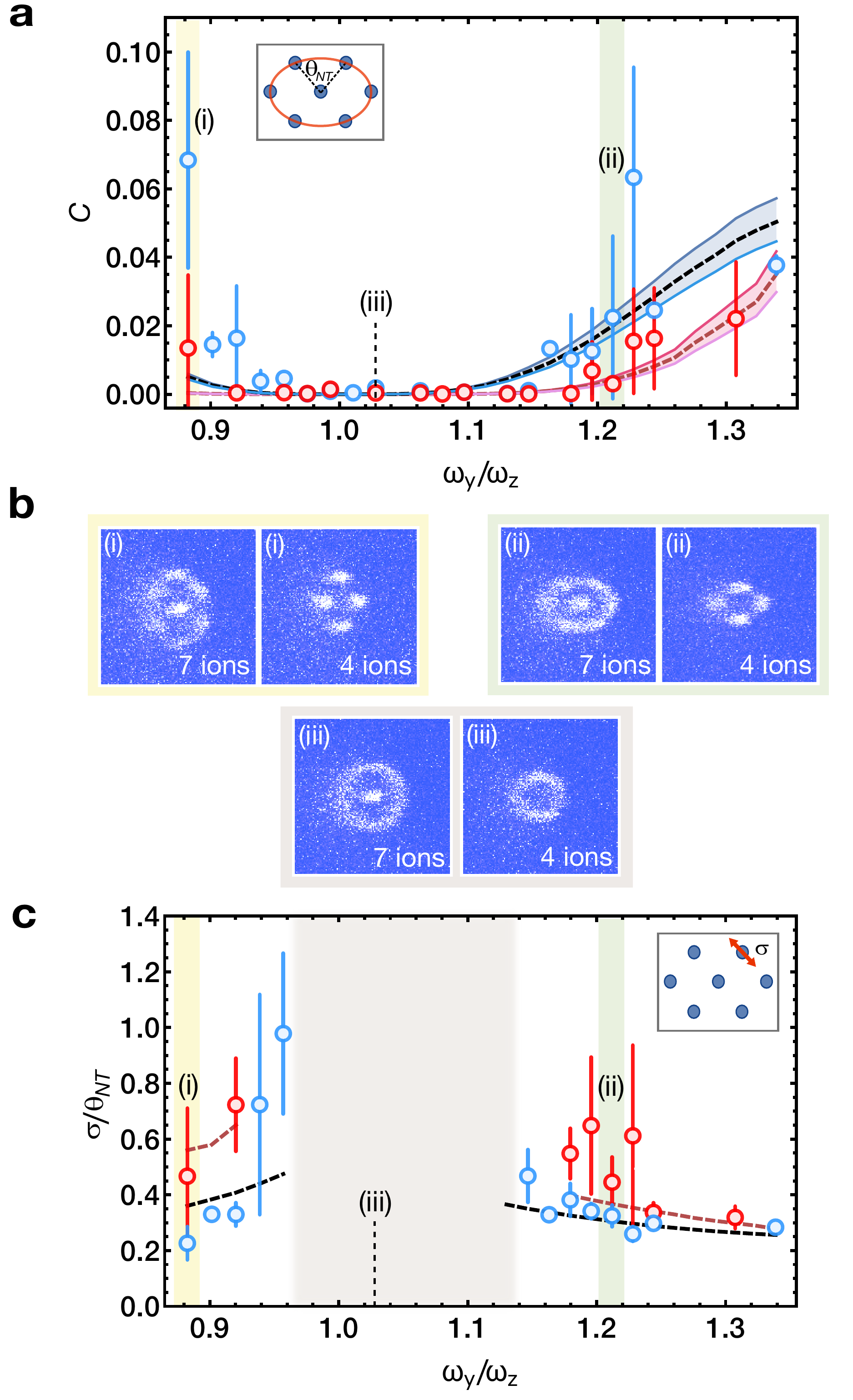}
    \caption{\textbf{Characterization of orientational melting for 4 and 7 ions.}
    \textbf{a}, Amplitude $C$ of the angular density-density correlation function $g(\theta_{NT})$ calculated along the elliptic trajectory shown in the inset (see text and Supplementary Materials). The onset of melting is clearly different for 4 (blue data) and 7 (red data) ions, as shown in the raw images in \textbf{b} taken in the regions (i) and (ii). The dashed black (red) line corresponds to the correlation amplitude expected for a crystal of 4 (7) ions at the best fitting temperature of $E_{T4}/k_B=102$~mK ($E_{T7}/k_B=96$~mK), as calculated from a Monte Carlo simulation (see Supplementary Materials). The blue (red) shaded area for 4 (7) ions represents a change of $\pm 10$~mK from the best fitting theory curve.
    \textbf{c}, Increase of the angular spread $\sigma$ as the melting transition is approached (see inset). $\sigma$ is obtained by fitting the density distribution (see Supplementary Materials), and the values are normalized by the angular separation $\theta_{NT}$ between the ions. The grey central area corresponds to trap conditions for which no density modulation is visible, as illustrated by image (iii) in \textbf{b}. The dashed black and red lines correspond to the theory curves from the simulation of 4 and 7 ions, respectively, at temperatures $E_{T4}/k_B$ and $E_{T7}/k_B$.
    The error bars in \textbf{a, c} indicate the standard deviation of the mean over 3 to 10 images.}
    \label{fig:3}
\end{figure}

\begin{figure*}[ht]
  \includegraphics[width=\textwidth]{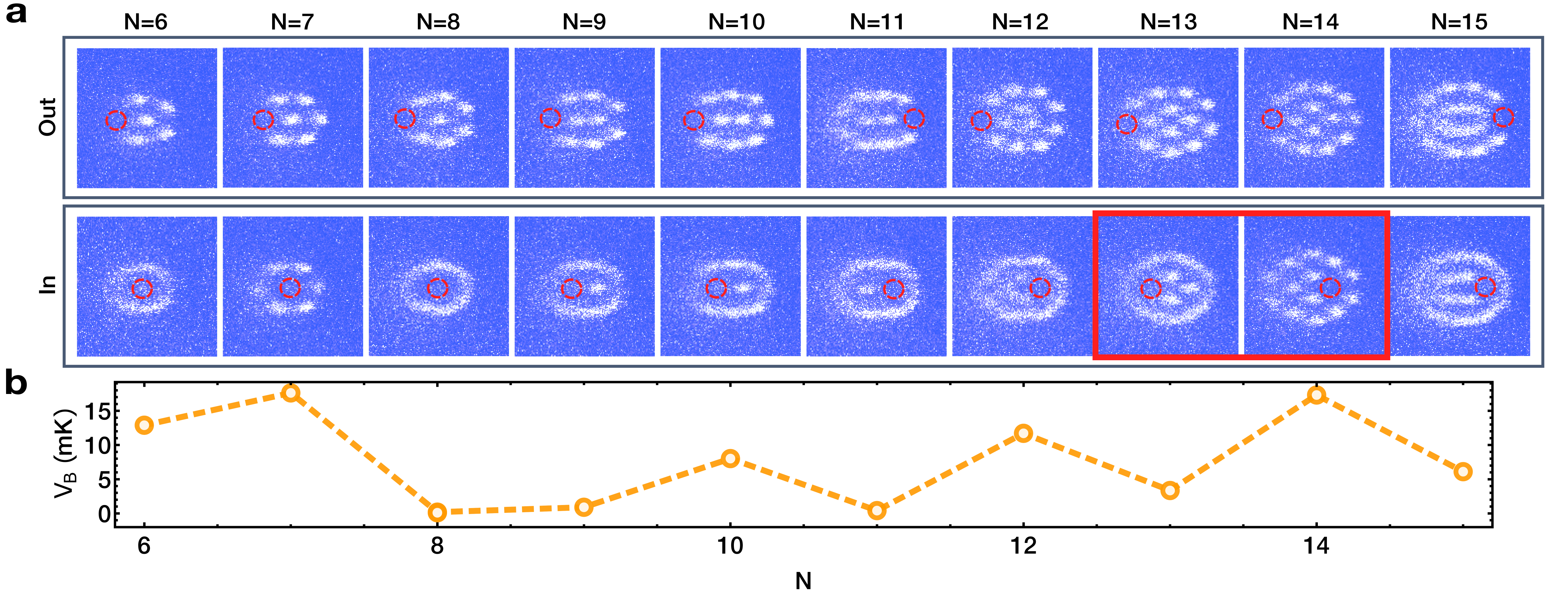}
  \caption{\textbf{Orientational melting in the presence of a pinning impurity.} \textbf{a}, Images of a crystal of $6$ to $15$ ions in the presence of an impurity ion of a different isotope. The data are taken at the onset of melting $\omega_y/\omega_z=1.18$ and $q_y=-0.182$, corresponding to $(\omega_x, \omega_y, \omega_z)=(401, 116, 98)$~kHz. The impurity appears as a dark ion (red circle), as it is not resonant to the cooling light. The upper (lower) images correspond to the impurity located in the outer (inner) shell. The impurity ion experiences a different trapping potential than the other ions. As a result, it suppresses melting in the hosting shell, and it is located typically along the trap $z-$axis (see Supplementary Materials). The images where the impurity is in the inner shell show that different shells can have independent behaviours. In particular, a crystal and a ring of delocalized particles can co-exist, see e.g. $N=13$. The strong dependence of the transition from the number of particles is shown by comparing the images in the red box ($N=13, 14$). Adding a single ion changes the periodicity of the density distribution and, therefore, the degree of localization. \textbf{b}, Height of the energy barrier for different ion numbers calculated with a Monte Carlo simulation. The barrier corresponds to the rigid rotation of the outer shell in the presence of a pinned inner shell (if present) (see Supplementary Materials). The cases of crystallized outer shells ($N=7,14$) correspond to the highest barriers.}
  \label{fig:4}
\end{figure*}

Fig.\ \ref{fig:3}\textbf{a} shows the amplitude of angular correlations $C$ measured for 4 and 7 $^{138}$Ba$^+$ ions as a function of $\omega_y/\omega_z$. The data show that the crystal loses and retrieves localization as $\omega_y/\omega_z$ is changed across 1. The change is continuous and the onset of melting is dependent on the ion number, as images (i) and (ii) in Fig.\ \ref{fig:3}\textbf{b} illustrate. The images provide a clear indication that the transition has no universal character.
The crystal melting is initiated by thermal fluctuations, as the ions' kinetic energy and the energy barrier $V_B$ become comparable. We compare the measured angular correlation $C$ with the results from a Monte Carlo simulation in which the ions' density distribution is calculated for different temperatures (see Supplementary Materials). The two curves that provide the best fit correspond to an angular kinetic energy of $E_{T4}/k_B=102$~mK and $E_{T7}/k_B=96$~mK for $4$ and $7$ ions, respectively. These values are comparable with the temperatures of Doppler cooled ion crystals in Paul traps with a similar geometry \cite{Feng2016,Kato2022}.

Additionally, we measure the angular spread $\sigma$ of the ion density distribution along the elliptical path by fitting the density profiles for $4$ and $7$ ions with a multi-gaussian function (see Supplementary Materials). We perform the fit only on the data for which the spatial modulation is non-negligible, corresponding to the values of $\omega_y/\omega_z$ for which $C>4\times 10^{-4}$. The data are plotted in Fig.\ \ref{fig:3}\textbf{c}. As expected, the angular spread increases as the crystal begins to melt. Additionally, we perform the same analysis on the simulated density profiles calculated from the Monte Carlo simulation at a kinetic energy $E_{T4}$ and $E_{T7}$ for $4$ and $7$ ions, respectively. The results from the simulation are in good agreement with the experimental data.  

\subsection{
Orientational melting in the presence of an impurity}

In order to directly observe the existence of magic numbers for which orientational melting is disfavoured, we increase the level of control over the melting transition by locally inducing the crystallization of a single shell with the use of a pinning impurity.

We realize this scenario, which was suggested in a similar fashion for electrons in a quantum dot \cite{GOLUBNYCHIY2003}, by deliberately adding one ion of a lighter isotope of Ba$^+$ into the crystal. We detect the presence of the impurity as a dark spot in the crystal, since the impurity is not resonant to the cooling light (see Supplementary Materials), as shown in Fig.\ \ref{fig:4}\textbf{a}.
A lighter isotope is more deeply trapped than $^{138}$Ba$^+$, and experiences a larger value of $\omega_y/\omega_z$ (see Supplementary Materials). As a result, the energy barrier for the rotation of the shell hosting the impurity is increased. Therefore, when $\omega_y/\omega_z$ is set at the crossover of the transition, the impurity inhibits melting in the shell where it is located, thus controlling, in practice, the degree of localization of all the ions in that shell.
Fig.\ \ref{fig:4}\textbf{a} shows the occurrence of melting in a crystal of $6$ to $15$ ions at the onset of the melting ($\omega_y/\omega_z=1.18$) and in the presence of one impurity. 
When the impurity is located in the inner shell, the outer shell can still undergo melting, see for example $N=13$ in Fig.\ \ref{fig:4}\textbf{a}. This is an evidence that the different shells can behave independently to one another while preserving the total angular momentum \cite{Schweigert1995, BONITZ2002}.
We exploit the local control of crystallization to reveal how the melting transition in a mesoscopic system strongly depends from the number of particles. In particular, we find that for our parameters melting is suppressed for $N=7$ and $N=14$, thus confirming the presence of "magic numbers" for which melting is disfavored \cite{Bedanov1994,Filinov2001}. The difference between the $N=13$ and $N=14$ cases is particularly striking, as adding just one particle in the system changes the whole collective behaviour.

We calculate the height of the energy barrier $V_B$ for the rigid rotation of the un-pinned shell around a pinned inner shell by using a Monte Carlo simulation (see Supplementary Materials). The results of the simulation are shown in Fig.\ \ref{fig:4}\textbf{b}. The trend of the energy barrier height is consistent with our observation: the barrier height $V_B$ is larger for the "magic numbers" $N=7$ and $N=14$ which correspond to the most stable crystalline configurations among the ion numbers we explored \cite{Schweigert1995, Belousov_1999}.

\subsection{Conclusions}


In conclusion, our results illustrate the direct observation of orientational melting in a 2D mesoscopic system of charged particles with repulsive Coulomb interactions. We observe evidence of the non-universality of the transition, and find excellent agreement with the results of a Monte Carlo simulation. Moreover, we use a single impurity to locally control the crystallization of the particles, and create new structures in which a crystal and a ring of delocalized particles co-exist.
Our system represents an ideal platform for studying thermodynamics in small systems in which the thermodynamic limit is not valid \cite{Hill1962,Jarzynski2011,Dauxois2002}. Interestingly, if the system is brought to sufficiently low temperatures (e.g. in a static trap \cite{Perego2020}), the role of quantum fluctuations prevails \cite{Bonitz2008B}, and fundamental quantum phenomena like the emergence of quantum statistics in a system of charged particles \cite{Roos2017} could be observed. Moreover, by achieving further control on the rotation of an individual shell, the control of multi-ring rotors at a quantum level could be achieved \cite{Urban2019}, with applications in sensing \cite{Campbell_2017} and in fundamental physics \cite{Horstmann2010}.
Finally, we note that in the presence of at least two shells and an impurity, the melting transition can also be interpreted as an effect of friction between the two independent concentric shells. This system could lead to a new approach for studying friction between two rotating periodically-rugged surfaces \cite{Hirano1990} with no edges \cite{Bylinskii_2016,Kiethe2017}. In this context, the number of ions in each shell could be additionally controlled by producing isomeric excitations of the crystal \cite{Bolton1993}. 














\section{Acknowledgments}
We gratefully thank Federico Berto, Roberto Concas, and Amelia Detti for their support in the realization of the experimental apparatus. We thank Guido Pupillo, Michael Drewsen, Chiara Menotti, Giovanna Morigi and the Quantum Gases group at LENS for fruitful discussions. Moreover, we are grateful to Martina Knoop, Boris Blinov, and Guenter Werth for helpful discussions during the assembly of the experimental setup.
This work was financially supported by the ERC Starting Grant PlusOne (Grant Agreement No. 639242), and the FARE-MIUR grant UltraCrystals (Grant No. R165JHRWR3).

\bibliography{rotor}

\clearpage

\section{Supplementary materials}
\subsection{The Paul trap}

\renewcommand{\thefigure}{S\arabic{figure}}
\setcounter{figure}{0}

\begin{figure*}[t]
    \centering
    \includegraphics[width=\textwidth]{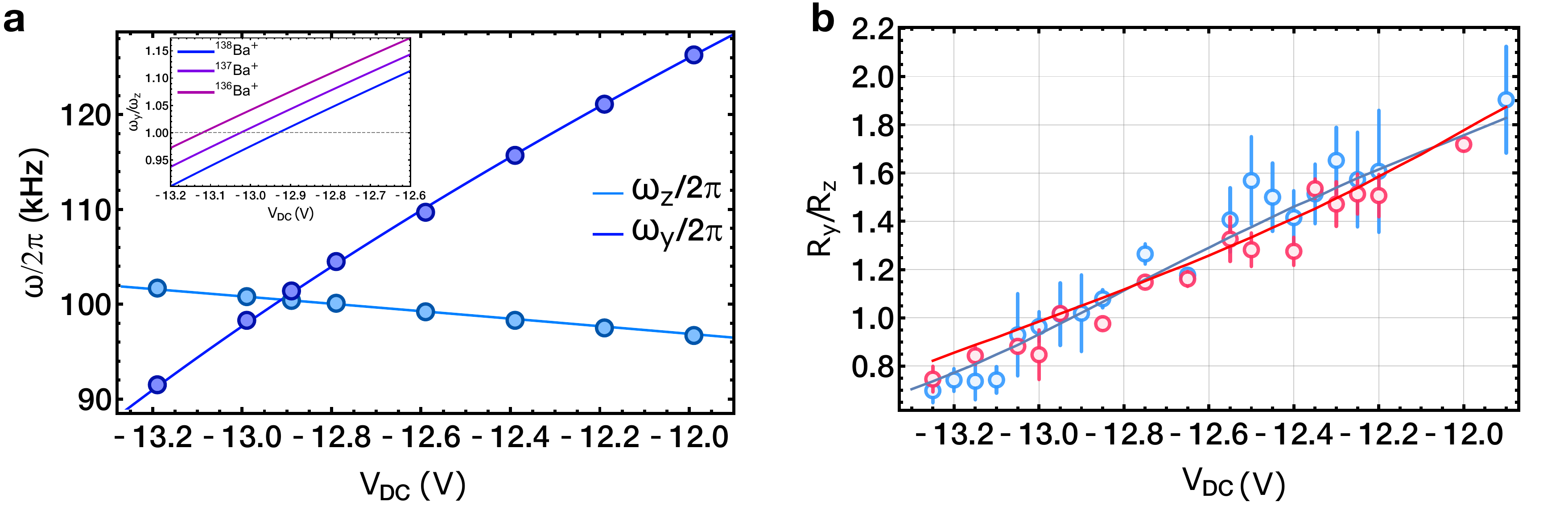}
\caption{\textbf{Trap frequencies calibration.} \textbf{a}, $\omega_y$ and $\omega_z$ trap frequencies measured via parametric heating. Data are plotted with fitting curves in the range of $V_{DC}$ relevant for the data shown in the main text. Errors are below 1\% therefore not shown. The symmetric trap is obtained for $V_{DC}=-12.9(1)V$. Inset shows $\omega_y/\omega_z$ calculated for different isotopes of Barium. Clearly, the symmetric case (dashed line), happens at different $V_{DC}$ for different Ba$^+$ isotopes. \textbf{b}, Aspect ratio $R_{y0}/R_{z0}$ of the ion cluster obtained from the elliptical fits of the ions' position for images with 4 (blue) and 7 (red) ions. The circular cluster is compatible with the $V_{DC}$ value from \textbf{a}: we extract -12.93(6)V and -12.93(7)V for 4 and 7 ions. Errors are standard deviations of the mean of 2 to 9 images. Solid lines are the values obtained from the Monte Carlo simulations of the ions' positions.}
    \label{supptrapf}
\end{figure*}

The linear Paul trap (Fig.\ \ref{fig:1}\textbf{a}) is composed by four electrodes fed with a radio-frequency (RF) voltage $V_{RF}$ at a frequency $\Omega_{RF} = 2\pi\times 4.7$~MHz providing confinement in the x-y plane, and by four DC electrodes fed with a voltage $V_{DC}$ providing confinement along the z axis, which is also the micromotion-free axis \cite{Perego2020}. 
In the adiabatic approximation, the time-averaged harmonic secular potential is parameterized by the trap frequencies $\omega_l=\frac{\Omega_{RF}}{2}\sqrt{a_l+\frac{q_l^2}{2}}$ with $l=(x,y,z)$, derived from the Mathieu parameters of our trap geometry $\mathbf{a},~\mathbf{q}$, which are, most generally, $\mathbf{a}=(a_x,a_y,a_z)$ and $\mathbf{q}=(q_x,q_y,0)$, with $q_y=-q_x$, $|\mathbf{a}|\propto V_{DC}/m$, $|\mathbf{q}|\propto V_{RF}/m$ and where $m$ is the particle mass \cite{Leibfried2003,Perego2020}. The value $q_y=-0.182$, used for extracting the data of Figs.\ \ref{fig:3} and \ref{fig:4}, corresponds to $V_{RF}=802~\text{V}_{pp}$. Due to the design shown in Fig.\ \ref{fig:1}\textbf{a}, the trapping potential is never three-dimensional in the whole stability diagram for the chosen ion numbers. This is due to the fact that the ratio $\omega_x/\omega_y$ in the perfectly round configuration is fixed at $4.1$ for any $\bf{q}$ parameter. As a result, the planar configuration would be lost with about 130 ions \cite{Richerme2016}, well above the number of ions we load. 

The micromotion is compensated for by minimizing photo-correlation signals \cite{Berkeland1998}, while trap frequencies are independently measured by parametric heating. 
The measured resonances relevant for the graphs of the main text are shown in Fig.\ \ref{supptrapf}\textbf{a} as a function of $V_{DC}$. From these data and their fit, we extract the trap parameters $a_l$ and $q_l$, as well as the trap frequencies of the other isotopes since $a_l\propto V_{DC}/m$ and $q_l\propto V_{RF}/m$ \cite{Perego2020}. We find $\omega_y/\omega_z=1$ at a different $V_{DC}$ for different isotopes, as shown in the inset plot of Fig.\ \ref{supptrapf}\textbf{a}. 

The trap frequency calibration is also cross-checked by calculating the ion crystal aspect ratio $R_{y0}/R_{z0}$ from the images with 4 and 7 ions used for extracting the data of Fig.\ \ref{fig:3}. The data in Fig.\ \ref{supptrapf}\textbf{b} are obtained from the ellipse fitting of the experimental images described in the Image analysis section. Data are fitted with a linear model (not shown) from which we extract the voltages corresponding to $R_{y0}/R_{z0}=1$: -12.93(6)~V and -12.93(7)~V for 4 ions and 7 ions, respectively. These values agree with the one obtained from the fits in Fig.\ \ref{supptrapf}\textbf{a} which is -12.9(1)V, corresponding to $\omega_y/\omega_z=1.00(4)$. The measured aspect ratios are also compared with the ones obtained by Monte Carlo simulations (solid lines in Fig.\ \ref{supptrapf}\textbf{b}), which agree with the data. The agreement between aspect ratio, simulations, and trap frequency calibration rules out large systematic errors on trap depth due to micromotion and any additional interparticle interaction terms in the relative equations of motion, not considered in our modelling. This correction is not observed within our uncertainties on trap frequencies, and we estimate it to be on the order $1$\% of $\omega_y/\omega_z$ by calculating the correction on the trap depth for 2 ions at $\mathbf{a}$ and $\mathbf{q}$ for which $\omega_y/\omega_z=1$ \cite{Moore1994}.

\subsection{Ions production and cooling}
$^{138}$Ba$^+$ ions are produced by a two-photon photoionization process with a 413~nm light that addresses the $^3D_{1}$ transition of neutral Barium \cite{Leschhorn2012} emitted by a resistive oven. The photoionization laser beam is orthogonal to the direction of the atom beam produced by the oven. Given that the smallest isotope shift of this transition is 101~MHz between $^{138}$Ba and $^{137}$Ba \cite{Dammalapati2009} and that our power broadening of the transition is around 30~MHz, the process has a small but non-negligible probability to photoionize other isotopes like $^{137}$Ba and, less probably, $^{136}$Ba. We post-select images with one or no dark ions for extracting the data shown in this paper. 

After production, $^{138}$Ba$^+$ ions are Doppler cooled by two orthogonal beams close to resonance with the $6S_{1/2}\rightarrow 6P_{1/2}$ transition, 
and repumped by addressing the $6P_{1/2}\leftrightarrow 5D_{3/2}$ transition. $^{137}$Ba$^+$, instead, is sympathetically cooled by the rest of the crystal \cite{Bowe1999}. The beams' waist at the ions position is about $300~\mu$m so that the Doppler cooling force is homogeneous across the whole crystal and there is no force imbalance that can initiate rotation.

\subsection{Imaging system}
The images of the ions are obtained by collecting their fluorescence with a first lens of numerical aperture NA=0.17 and a magnification setup of x20. The ions' fluorescence is integrated for 1~s by the camera to obtain the spatial distribution of photons emitted by each ion. As a result, images need to be considered as probability density plots rather than as images of individual particles. Additionally, a similar imaging setup is used to monitor the ions' florescence with a photon counter to count the ion number during the experimental sequence, even in the melted phase.

\subsection{Image analysis}
\begin{figure*}[t]
    \centering
    \includegraphics[width=110mm]{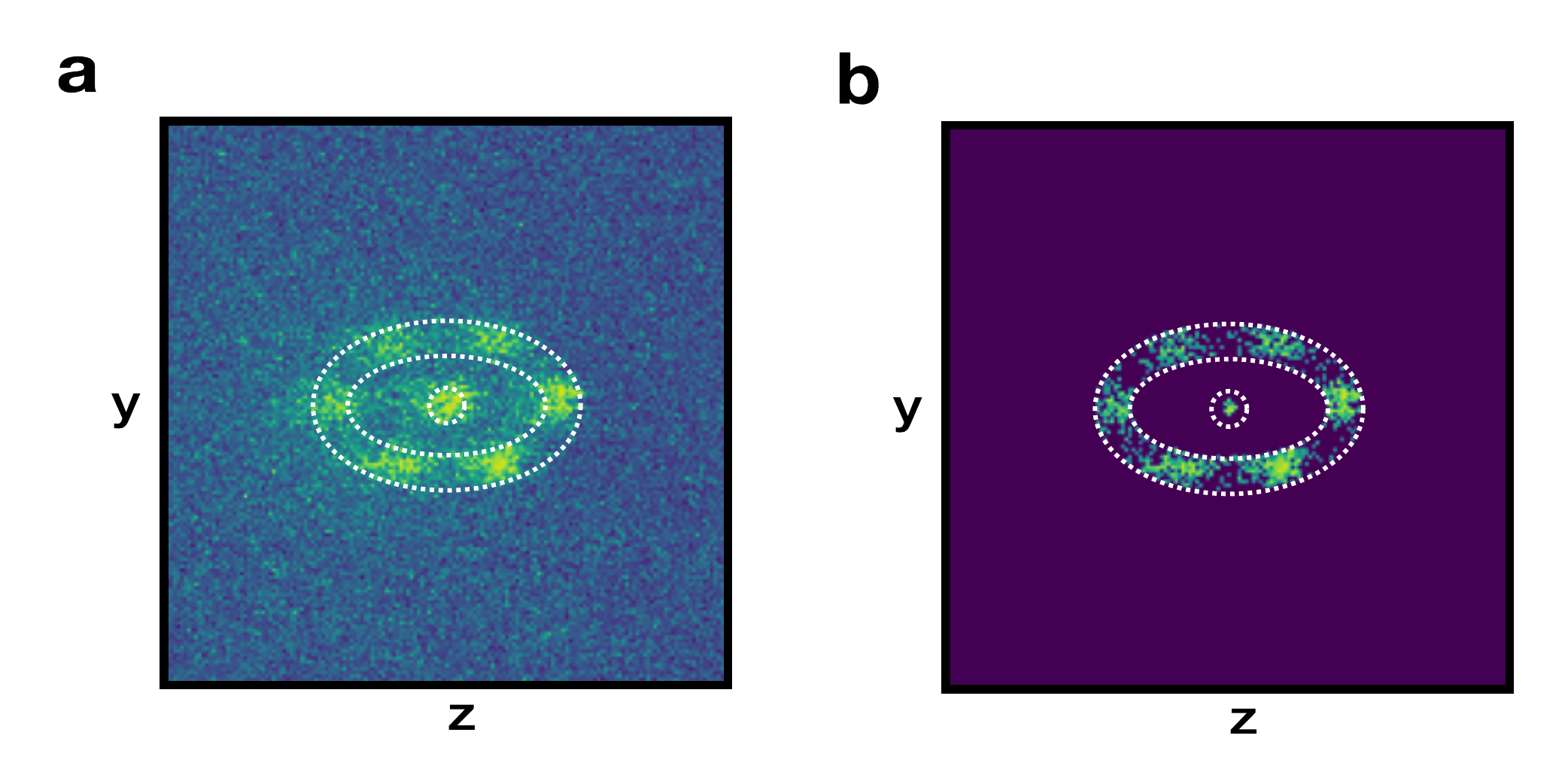}
    \caption{\textbf{Trajectory selection.} \textbf{a}, The trajectory describing the ions' positions is found by maximizing the photon counts of the $4\delta^{2}\times N$ brightest pixels inside an elliptical ROI. In \textbf{b}, the pixels that are considered are shown.
    }
    \label{trajectory}
\end{figure*}

The ions' arrangement in a certain shell has, most generally, an elliptical shape. Therefore, we convert the images to elliptic coordinates to correctly evaluate and compare angular correlations and particle's spread at different $\omega_y/\omega_z$. We define the relationship between the Cartesian coordinates $(z,y)$ and the elliptic coordinates $(r,\theta)$ corresponding to the De La Hire representation as
\begin{equation}
z = R_{z} r \cos \theta
\end{equation}
\begin{equation}
y = R_{y} r \sin \theta,
\end{equation}
where
\begin{equation}
R_{z} = \frac{\sqrt{2}R_{z0}}{\sqrt{R_{z0}^{2} + R_{y0}^{2}}}
\end{equation}
\begin{equation}
R_{y} = \frac{\sqrt{2}R_{y0}}{\sqrt{R_{z0}^{2} + R_{y0}^{2}}}.
\end{equation}
$R_{z0}$ and $R_{y0}$ are the two semi-axes along z and y of the ellipse that encloses the positions of the ions in a certain shell of the cluster.
To find the ellipse that best encloses the ions in an image, and to analyze the data, we must define an elliptical region of interest (ROI), see Fig.\ \ref{trajectory}. An elliptical ROI is formed by the points with coordinates $(z,y)$ that satisfy the condition:
\begin{equation}
\frac{(z-O_{z})^{2}}{(R_{z0} + \delta)^{2}} + \frac{(y-O_{y})^{2}}{(R_{y0} + \delta)^{2}} < 1 < \frac{(z-O_{z})^{2}}{(R_{z0} - \delta)^{2}} + \frac{(y-O_{y})^{2}}{(R_{y0} - \delta)^{2}},
\end{equation}
where $(O_{z}, O_{y})$ are the coordinates of the center of the crystal and $\delta$ is the half width of the ROI. 
In case the crystal structure has an ion at the center, we also define a circular ROI$_c$ that encloses it. This is defined by the condition:
\begin{equation}
\sqrt{(z - O_{z})^{2} + (y - O_{y})^{2}} < \delta.
\end{equation}

In the image analysis, we first subtract from each image a background image taken with the lasers on but without ions. Then, we identify the parameters $O_{z}$, $O_{y}$, $R_{z0}$, and $R_{y0}$ corresponding to the ellipse that better inscribe the ions' positions. To this end, we define $\delta=5$ and find the optimal parameters of the ROI for which the total number of photon counts in the $4\delta^{2}N$ brightest pixels enclosed in the ROI is maximized. We use these parameters to extract the aspect ratio of the ion crystal $R_{y0}/R_{z0}$. The yellow data in Fig.\ \ref{fig:2} correspond to the trap parameters for which $R_{y0}=R_{z0}$.

\begin{figure*}[ht]
    \centering
    \includegraphics[width=\textwidth]{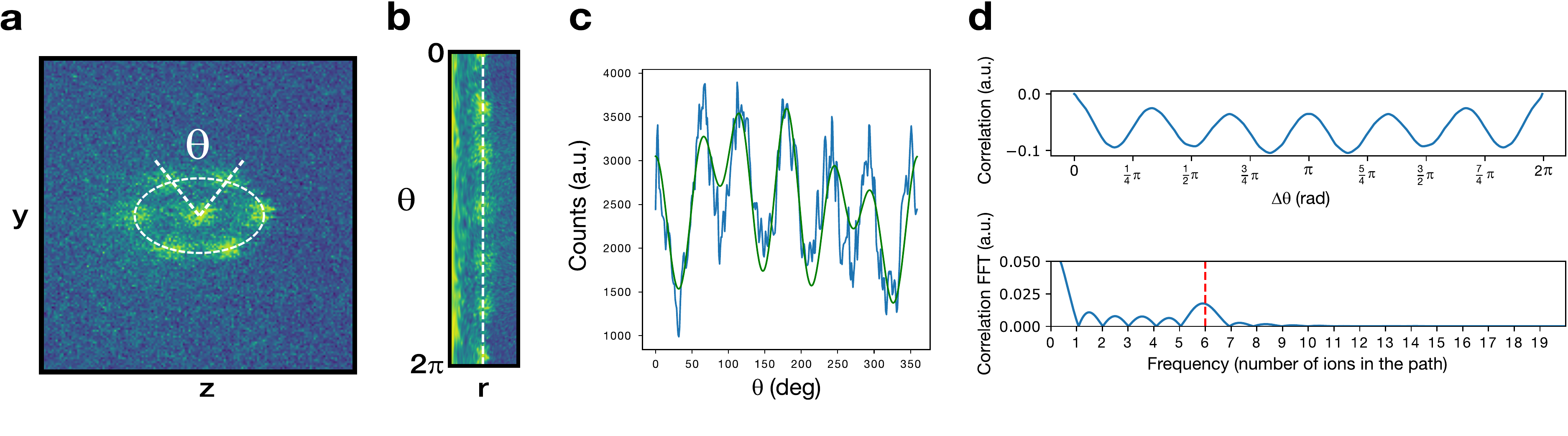}
    \caption{\textbf{Image analysis and correlation measurement.} \textbf{a}, An image of a crystal in Cartesian coordinates. The ions' positions in a certain shell are parameterized by the elliptical path shown in figure (dashed ellipse). \textbf{b}, Same image in elliptical cylindrical coordinates $(r, \theta)$ shows the periodic structure of the shell with the ellipse transformed into a line (dashed line). \textbf{c}, Photon counts after integrating the pixel counts in \textbf{b} along $r$. A modulation along the trajectory can be seen when the particles form a crystal. The multi-gaussian fit of the peaks provides the spread $\sigma$ of the angular distribution of the individual ions. \textbf{d} Top, angular correlations calculated by applying the formula in eq.\ \ref{gSupplementary} to the data in \textbf{c}. In case the ions form an ordered crystal structure, the correlation function shows a modulation (upper plot). We extract the amplitude of this modulation from the Fourier transform of the correlation function (bottom plot), which shows a peak at the periodicity corresponding to the number of particles $N_T$ in the shell under consideration, 6 here (red line). In the melting phase no modulation is visible.
    }
    \label{corr}
\end{figure*}

After finding the elliptic trajectory enclosing the ions' positions (see Fig.\ \ref{corr}\textbf{a}), we project the image from cartesian coordinates to elliptical coordinates, see Fig.\ \ref{corr}\textbf{b}. The conversion of the images to elliptic coordinates is done with a bilinear interpolation method to correctly convert the pixels values in the new frame.
Once the image has been transformed in elliptical coordinates, we select a single shell by considering an area of width 40 pixels enclosing the shell in the radial direction $r$. In order to obtain the angular density $n(\theta)$ we integrate the 40-pixels-wide area along $r$, see Fig.\ \ref{corr}\textbf{c}. 
The ions' spread plotted in Fig.\ \ref{fig:3}\textbf{c} is obtained by fitting $n(\theta)$ with a multi-gaussian fit from which we extract the standard deviation $\sigma$ representing the angular spread of the particles, see Fig.\ \ref{corr}\textbf{c}. In the fits, we impose that all gaussians have the same $\sigma$, and that they are centered around the positions expected for a zero-temperature crystal (e.g. for 4 ions $\theta=0,\pi/2,\pi,3\pi/2$). 

The angular correlation function calculated to extract the data of Fig.\ \ref{fig:3}\textbf{a} is defined as
\begin{equation}
g(\Delta \theta) = \frac{\sum_{\theta = 0}^{2 \pi} n(\theta) n(\theta + \Delta \theta) - \sum_{\theta = 0}^{2 \pi} n(\theta)^{2}}{\sum_{\theta = 0}^{2 \pi} n(\theta)^{2}},
\label{gSupplementary}
\end{equation}
which describes the probability of finding an ion at an angle $\Delta \theta$ relative to an other particle along the elliptical path parameterised by $\theta$. 
When the ions are localized, $g(\Delta\theta)$ has a non-negligible periodic modulation with a periodicity that depends on the number of particles in the shell under consideration (see Fig.\ \ref{corr}\textbf{d}). In order to find the amplitude of this modulation, we first perform the Fourier transform of $g(\Delta\theta)$. The Fourier transformed data present a peak at the frequency corresponding to the angular spacing $\theta_{NT}$ between the ions inside the shell, e.g. $\theta_{NT}=\pi/2$ for 4 ions. The amplitude $C$ of this peak is reported for different values of $\omega_y/\omega_z$ in Fig. \ \ref{fig:3}\textbf{c}.


\subsection{Monte Carlo simulations}

\begin{figure*}[t]
    \centering
    \includegraphics[width=110mm]{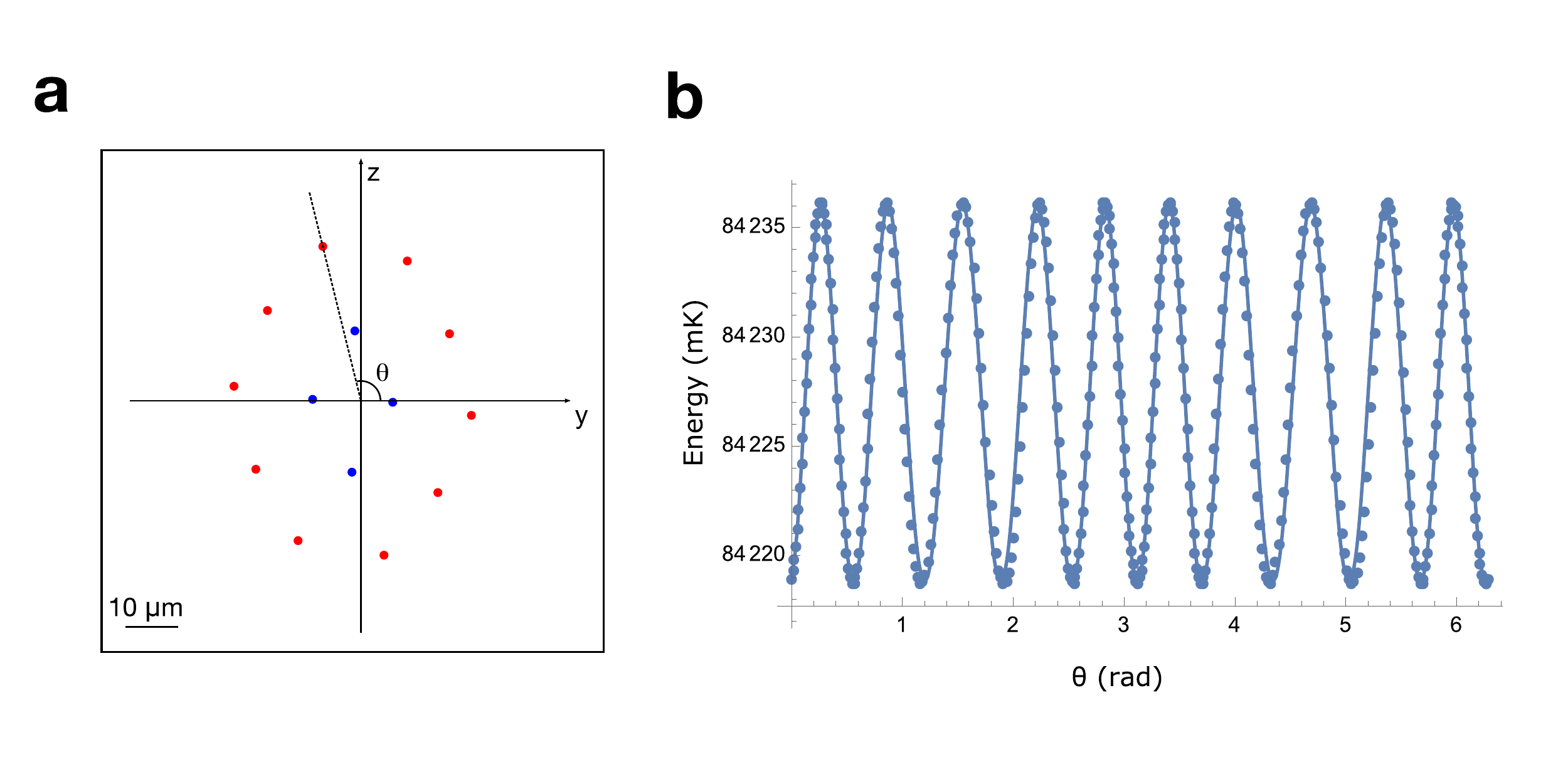}
    \caption{\textbf{Monte Carlo simulation.} \textbf{a}, Ground state configuration found for a crystal of $N=14$ ions in which the outer shell (red points) is rotated by an angle $\theta$. The ions in the inner shell (blue points) are kept fixed in the absolute ground state position, which was calculated separately, while the other ions are free to move in the plane. In order to simulate the rigid rotation of the outer shell, one of the free-to-move ions is bounded to a motion on a line at an angle $\theta$ from the absolute ground state configuration.
    \textbf{b}, plot of the energies of the system as a function of $\theta$ as found from the Monte Carlo simulation (blue points). The fitting function (blue line) is $V(\theta)$.
    }
    \label{barrier}
\end{figure*}

We find the equilibrium position of the ions by using a Monte Carlo simulation. In this simulation, each ion is initially placed stochastically inside an area of size $100~\mu$m$\times100~\mu$m, which is divided in unit cells of size $25$~nm$\times25$~nm. In order to find the positions of the ions that minimize the total energy, we displace one ion at a time within an area of $81\times81$ unit cells, and calculate the total energy for each position. After this calculation, the ion position is changed to the one that minimizes the energy. Afterwards, a new ion is randomly chosen and the energy minimization protocol is repeated. If the energy is not minimized after $10\times N$ steps in a row, the energy minimization protocol is repeated one last time for all the ions. If no other minimum in energy is found, the procedure is stopped.
In order to make sure that we are not lying in a local minima, the whole procedure is repeated at least 5 times and the ions' positions minimizing the energy are assumed to constitute the ground state of the crystal configuration.

In order to estimate the height of the energy barrier $V_B$, we calculate the energy associated to a rigid rotation of the crystal. To this end, we repeat the same Monte Carlo simulation, but we impose that one ion can only move along a line at an angle $\theta$ from the $z-$axis (see Fig.\ \ref{barrier}\textbf{a}). In case of a multi-shell simulation as for the data of Fig.\ref{fig:4}\textbf{b}, the position of the ions in the inner shell are not varied, and the energy of the crystal is minimized by moving the ions of the outer shell, with one ion moving along the line at an angle $\theta$. 

%
The potential $V(\theta)$ associated to the rigid rotation of the crystal has, for $\omega_y\simeq\omega_z$, a sinusoidal shape with amplitude $V_B/2$ (see Fig.\ \ref{barrier}\textbf{b}) \cite{Schweigert1995}. However, when the ion crystal has a larger ellipticity, $V(\theta)$ is not longer well fitted by a sinusoidal function. For this reason, we perform a change of variable through the transformation $\tan{(\theta)}=\eta \tan{(\theta_E)}$ and adapt the parameter $\eta$ in order to make the data periodic in $\theta_E$. Then, we use a power expansion of the cosine to find the potential $V(\theta_E)$:
\begin{equation}
\begin{split}
V(\theta_{E}) &= a + b \cos (N_{T} \theta_{E} + \phi) + c \cos^{3} (N_{T} \theta_{E} + \phi) \\
&+ d \cos^{4} (N_{T} \theta_{E} + \phi) + e \cos^{5} (N_{T} \theta_{E} + \phi)
\end{split}
\end{equation}
where $N_{T}$ is the number of ions in the shell.

After that we model $V(\theta_E)$ for all the values of $\omega_y/\omega_z$ that we consider, we derive the angular density of the particles by using a Boltzmann thermal distribution at a temperature $T$, and then change the angular variable back to $\theta$. In defining the angular density, we include the effects of the spread due to the limited resolution of the imaging system. 

Once the densities are extracted, we perform the same procedure as for the data in Figs.\ref{fig:3}\textbf{a} and \textbf{c} to derive the theoretical model. In particular, we calculate $C$ in Fig. \ref{fig:3}\textbf{a} for different temperatures from $T=1$~mK to $T=120$~mK at steps of $T=1$~mK. We then perform a least squared error analysis to find the curves that better describe the data, corresponding to $E_{T4}/k_B=102$~mK and $E_{T7}/k_B=96$~mK for $4$ and $7$ ions, respectively.
We perform the same analysis for the $7$ ions crystal of Fig.\ \ref{fig:4}\textbf{a} with the impurity ion at the center (bottom panel). We find a temperature that is lower than $E_{T7}$ by a factor of $3$. We attribute this difference to an improvement of the high voltage amplifiers feeding the DC electrodes that was implemented after that the data in Fig.\ref{fig:3} were taken.

\end{document}